\documentstyle[preprint,prb,aps]{revtex}

\tighten

\begin{document}
\preprint{\em Submitted to Physica C}

\draft

\title{Proximity and Josephson effects in superconductor - two
dimensional electron gas planar junctions.}

\author{A.~F. Volkov}

\address{Institute of Radioengineering \& Electronics of the Russian
Academy of Sciences,\\ Mokhovaja 11, Moscow 103907, Russia}

\author{P.~H.~C.~Magn\'ee,  B.~J.~van Wees, and T.~M.~Klapwijk}

\address{Department of Applied Physics and Material Science Center,
University of Groningen,\\ Nijenborgh 4, 9747 AG Groningen, The
Netherlands.}


\maketitle

\begin{abstract}
The DC Josephson effect is theoretically studied in a planar junction
in which a two dimensional electron gas (2DEG) infinite in lateral
directions is in contact with two superconducting electrodes placed on
top of the 2DEG. An energy gap in the excitation spectrum is created in
the 2DEG due to the proximity effect. It is shown that under certain
conditions, the region of the 2DEG underneath the superconductors is
analogous to a superconducting region with an order parameter
$\varepsilon_g\exp(i\phi)$\/, where $\varepsilon_g~(\varepsilon_g<\Delta)$\/
depends on the interface transmittance and the Fermi velocity mismatch
between the superconductors and the 2DEG.
\end{abstract}
\vspace\baselineskip
\pacs{Key words: energy gap, Josephson effect, proximity effect,
two-dimensional electron gas, weak links}

\narrowtext

\section{Introduction}

In recent years significant progress has been made in the preparation
and study of Josephson junctions, in which the weak coupling is
realized through a two dimensional electron gas (2DEG). These systems
are prepared using semiconductor heterostructures with superconducting
contacts. The structure Nb-InAs(2DEG)-Nb appears to be the most
promising \cite{NATA92,NKH92,DHvWetal94}, although it may be possible
to obtain the Josephson effect also in other structures based, for
example, on GaAs \cite{Letal93,Getal93,MWA94}. In Josephson junctions
with a 2DEG as a weak link, one expects phenomena analogous to the
conductance quantization in quantum point contacts
\cite{vWetal88,Wetal88,BvH91a}. One of these phenomena is the
predicted quantization of the critical Josephson current $I_c$\/
\cite{BvH91b,FTT91}. In short quantum contacts ($L<\xi,~\xi=\hbar
v_{\text{F}}/\pi\Delta$\/ is the coherence length), the critical
current is predicted to increase with increasing contact width in a
step like way, with a step height equal to $e\Delta/\hbar$\/.

The geometry of the contacts shown in Fig.~1a has been considered in
previous theoretical works \cite{BvH91b,FTT91}. In this case, electrons
move in a channel or 2D region of finite length and experience Andreev
reflections at the SN interfaces. The excitation spectrum of the 2DEG (or
1DEG) is changed due to the interference of waves reflecting at the
opposite SN boundaries; in particular, bound states decaying into the S
regions and corresponding to energies $\varepsilon<\Delta$\/ appear in the
system \cite{Kul70,Ish70,BJ72}. These bound states give the main
contribution to the Josephson current $I_c$\/ at low temperatures.

It is, because of its practical significance, of interest to study the
DC Josephson effect in the geometry shown in Fig.~1b. In this
geometry, the 2DEG is unbounded in the $(x,y)$\/ plane (the $z$\/ axis
is directed perpendicular to the 2DEG plane). We will find the
condensate Green's functions in the N layer, i.e., in the layer with
the 2DEG, and show that due to the proximity effect, the properties of
the N layer at $|x|>L/2$\/, where $L$\/ is the distance between both
superconductors, are analogous to those of a superconductor with an
effective order parameter $\varepsilon_g\exp(i\phi)$\/ with
$\varepsilon_g$\/ dependent on the interface transmittance and the
Fermi velocity mismatch. We note that the pair potential in the N
layer $\Delta_{\text{N}}$\/ determined from the selfconsistency
equation equals zero because the electron-phonon coupling constant in
the N layer is supposed to be vanishing. For a general discussion of
the proximity effect in terms of a tunneling model see
Ref.~\onlinecite{McM68}. We adopt the following simplified model of
the system shown in Fig.~1b. The transmittance of the SN interface
$T_b(x)$\/ is proposed to be dependent on the coordinate $x$\/; it
varies from a given value $T_b$\/ at $|x|>L/2$\/ to zero at
$|x|<L/2$\/ on a characteristic length $x_0$\/. We assume that $x_0$\/
is much larger than the Fermi wave length $k_{\text{F}}^{-1}$\/ in the
N layer, but is smaller than the coherence length in the 2DEG
$\xi_{\text{2D}}=\hbar v_{\text{F}}/\pi\varepsilon_g$\/. Then, we can
use an adiabatic approximation to calculate the energy spectrum of the
2DEG at $|x|>L/2$\/.

\section{The proximity effect in the 2DEG.}

Let us consider the energy-diagram for a system shown schematically in
Fig.~2. The region at $z<0$\/ is occupied by a superconductor, and there is
a quantum well with the 2DEG in the layer $0<z<d$\/. The Fermi momenta in
the S and N regions ($p_{\text{F}}=\sqrt{p_{\parallel}^2+p_z^2}$\/ and
$k_{\text{F}}=\sqrt{k_x^2+k_y^2}$\/, respectively) differ greatly from
each other due to a significant difference in electron concentrations
in these regions; namely, $p_{\text{F}}=(3\pi^2n_{\text{S}})^{1/3}\gg
k_{\text{F}}=(2\pi n_{\text{2D}})^{1/2}$\/ (the formula for
$k_{\text{F}}$\/ is written for the case when only the lowest subband
in the quantum well is filled). Generally speaking, there exists a
Schottky barrier at $z=0$\/, which we model by a potential of the form
$U_ba_b\delta(z)$\/, where $U_b$\/ and $a_b$\/ are the height and
width of the barrier varying smoothly at $|x|\cong L/2$\/.

In order to calculate the excitation spectrum and wave functions of
the 2DEG, we write the well known Bogoliubov-de Gennes equations
\begin{equation}
\hat{H}\hat{\Psi}_k=\varepsilon_k\hat{\Psi}_k.
\label{BdG}
\end{equation}
\noindent Here $\hat{\Psi}_k$\/ is the two component wave function,
$\varepsilon_k$\/ is the excitation energy, and $\hat{H}$\/ is the
Hamiltonian of the system,
\begin{eqnarray}
\hat{H}&=&\left[\left(-\frac{1}{2m_1}\nabla^2 + U_ba_b\delta(z) -
\varepsilon_{\text{F}}\right)\hat{\sigma}_z +
\hat{\sigma}_x\Delta\right]\theta(-z)\nonumber\\
&+&\left[-\frac{1}{2m_2}\nabla^2 + U_{\text{Sm}} -
\varepsilon_{\text{F}}\right]\hat{\sigma}_z\theta(z)\theta(d-z).
\label{hamiltonian}
\end{eqnarray}
\noindent Here $m_1$\/ and $m_2$\/ are the effective masses in the S and N
regions respectively, and $U_{\text{Sm}}$\/ is the difference between the
potentials of the conduction band edges in the S and N regions (see
Fig.~2).

One can show that propagating (into the S region) states with
$\varepsilon_k>\Delta$\/ and bound states with $\varepsilon_k<\Delta$\/
exist. The latter states correspond to a branch of the spectrum with an
effective energy gap $\varepsilon_g$\/, which is small if the barrier
transmittance is small. For low temperatures ($T\ll\Delta$\/), the main
contribution to the critical current $I_c$\/ originates from these bound
modes. We will restrict ourselves to the case of low temperatures. A
solution describing the bound states has the from
\begin{eqnarray}
\hat{\Psi}_{k_{\parallel}}(r)&=&\exp(ik_{\parallel}r_{\parallel})\left
\{\left[B_+\exp(\kappa z+ip_zz)\left(\begin{array}{c}
                        u\\
                        v\end{array}\right) - B_-\exp(\kappa
                        z-ip_zz)
                      \left(\begin{array}{c}
                              v\\
                              u\end{array}\right)\right]\theta(-z)
\right.\nonumber\\
&+&\left[E_+\exp(ik_zz)\left(\begin{array}{c}
                             1\\
                             0\end{array}\right) -
         E_-\exp(-ik_zz)\left(\begin{array}{c}
                             1\\
                             0\end{array}\right) +
         H_+\exp(i\bar{k}_zz)\left(\begin{array}{c}
                                 0\\
                                 1\end{array}\right)\right.\nonumber\\
& &\left.-\left.H_-\exp(-i\bar{k}_zz)\left(\begin{array}{c}
                                  0\\
                                  1\end{array}\right)
\right]\theta(z)\theta(d-z)\right\}.
\label{wavefunc}
\end{eqnarray}

Consider first the wave function in the quantum well (the second term
in Eq.(\ref{wavefunc})). The first two terms ($E_{\pm}$\/) describe an
electron excitation moving forward and backward and the second
term ($H_{\pm}$\/) correspond to a hole excitation. Momenta of
these excitations along the $z$\/ axis can be presented in the form
$k_z=k_n+\delta k_n$\/ and $\bar{k}_z=k_n+\delta\bar{k}_n$\/, where
$k_n=(\pi/d)(n+1)$\/ and $|\delta k_n,\delta\bar{k}_n|\ll k_n$\/. Here
we will consider the case when only the lowest subband, with an energy
$\varepsilon_0=\pi/d^2m_2$, is filled ($n=0$\/) and other subbands
($n\geq 1$\/) are empty. So we are interested in the states with
$n=0$\/. A relation between $\delta k_0$\/ and $\delta\bar{k}_0$\/ can
be obtained from Eq.(\ref{BdG}),
\begin{equation}
\varepsilon_k = \xi_k+v_0\delta k_0 = -(\xi_k+v_0\delta\bar{k}_0).
\label{deltak}
\end{equation}
\noindent Here $\xi_k=(k_{\parallel}^2-k_{\text{F0}}^2)/2m_2$\/ is the
kinetic energy of electrons in the 2DEG relative to the Fermi energy,
$\varepsilon_{\text{F}} = (k_0^2+k_{\text{F0}}^2)/2m_2 +
U_{\text{Sm}}$\/, $k_{\text{F0}}$\/ is the Fermi momentum in the limit
of infinite barrier height, $v_0=k_0/m_2$\/.

Consider now the wave functions in the S region, which decay over the
length $\kappa^{-1}$\/. The coefficients $(u,v)$\/ have the usual form,
\begin{equation}
u^2 = \frac{1}{2}(1+\xi_p/\varepsilon_p),~v^2 =
\frac{1}{2}(1-\xi_p/\varepsilon_p),
\label{UandV}
\end{equation}
\noindent but in this case the functions $\xi_p$\/ are purely imaginary
\begin{equation}
\xi_p = -i\kappa p_z/m_1,~\varepsilon_p = (\Delta^2+\xi_p^2)^{1/2}.
\label{xiP}
\end{equation}
\noindent The momentum $p_z$\/ approximately coincides with the Fermi
momentum $p_{\text{F}}$\/ where we assume the reflection at the
interface to be specular, i.e. $p_{\parallel}=k_{\parallel}$\/ and,
as noted above, $k_{\parallel}\ll p_{\text{F}}$\/. Hence for the Fermi
energy, we can write
\begin{equation}
\varepsilon_{\text{F}} = p_{\text{F}}^2/2m_1 =
(p_z^2+p_{\parallel}^2)/2m_1 \cong p_z^2/2m_1.
\label{FermiE}
\end{equation}

In order to find the excitation spectrum, i.e. the dependence of
$\varepsilon_k$\/ on the momenta in the $(x,y)$\/ plane
$k_{\parallel}$\/, we must use boundary conditions at the interface
($z=0$\/). These conditions consist in continuity of
$\hat{\Psi}_{k_{\parallel}}(z)$\/ at $z=0$\/ and in a relationship
between the derivatives $\partial_z\hat{\Psi}_{k_{\parallel}}(z)$\/ at
$z=0$\/ \cite{MB84}
\begin{equation}
\hat{\Psi}_{k_{\parallel}}(+0) = \hat{\Psi}_{k_{\parallel}}(-0),
{}~\frac{1}{2m_2}\partial_z\hat{\Psi}_{k_{\parallel}}(z)|_{+0} -
\frac{1}{2m_1}\partial_z\hat{\Psi}_{k_{\parallel}}(z)|_{-0} =
U_ba_b\hat{\Psi}_{k_{\parallel}}(0).
\label{boundcond}
\end{equation}
\noindent Substituting Eq.(\ref{wavefunc}) into Eq.(\ref{boundcond}),
we get a set of algebraic equations for the coefficients $E_{\pm}$\/,
$H_{\pm}$\/ and $B_{\pm}$\/. The solvability condition results in a
dispersion relation for the excitation spectrum at
$\varepsilon_k<\Delta$\/
\begin{equation}
1 + \alpha_k\bar{\alpha}_k(w^2+s^2) + w(\alpha_k +
\bar{\alpha}_k) + is(\varepsilon_k/\xi_p)(\alpha_k -
\bar{\alpha}_k) = 0
\label{dispersion1}
\end{equation}
\noindent Here $\alpha_k=\delta k_0 d\ll 1$\/ and $\bar{\alpha}_k =
\delta\bar{k}_0 d= -(\alpha_k+2\xi_k/\varepsilon_0)$\/. $w =
(2U_ba_bm_2/k_0)$\/ is a dimensionless parameter characterizing the
barrier transmittance, the factor $s=p_{\text{F}} m_2/k_0 m_1$\/
depends on the mismatch of the Fermi momenta and the effective masses.
Eq.(\ref{dispersion1}) determines the spectrum of bound states with
energies $\varepsilon_k<\Delta$\/. Eq.(\ref{dispersion1}) can be
rewritten in the form
\begin{equation}
\varepsilon_k^2 \left[1 + \frac{2\varepsilon_{g0}}{(\Delta^2 -
\varepsilon_k^2)^{1/2}}\right] = (\xi_k -
\frac{w}{s}\varepsilon_{g0})^2 + \varepsilon_{g0}^2,
\label{dispersion2}
\end{equation}
\noindent where $(\xi_k-\frac{w}{s}\varepsilon_{g0})$\/ is the
relative kinetic energy of electrons moving in the $(x,y)$\/ plane.
The quantity $\varepsilon_{g0} = \varepsilon_0\; s/(w^2+s^2)$\/ is the
energy gap in the excitation spectrum of the 2DEG induced by the
proximity effect in the case of very low barrier transmittances.
Indeed, under the condition
\begin{equation}
(\varepsilon_0/\Delta)\frac{s}{w^2+s^2}\ll 1
\label{limitcond}
\end{equation}
\noindent one can neglect the second term in the square brackets in
Eq.(\ref{dispersion2}) and obtain for $\varepsilon_k$\/ not too close
to $\Delta$\/
\begin{equation}
\varepsilon_k = \pm \left[\varepsilon_{g0}^2 +
(\xi_k-\frac{w}{s}\varepsilon_{g0})^2\right]^{1/2}.
\label{lowTrlimit}
\end{equation}
\noindent Therefore, the dependence of $\varepsilon_k$\/ on
$k_{\parallel}$\/ is nearly the same as in a 2D superconductor with
the energy gap
\begin{equation}
\eqnum{12'}
\varepsilon_g\approx\varepsilon_{g0} = \varepsilon_0\frac{s}{w^2+s^2},
\label{lowTrEg}
\end{equation}
\noindent where $\varepsilon_0$\/ is the subband energy for $n=0$\/.
This dependence is shown in Fig.~3 for several values of $w$\/. In
Fig.~3 it is shown that, with changing $w$\/, not only the value of
$\varepsilon_g$\/ is changed, but also the $k_{\parallel}$\/ at which
the minimum in $\varepsilon_k$\/ occurs. This may be understood
intuitively, since increasing the SN-barrier transparency will alter
the exact form of the electron-hole wave function in the 2DEG,
$\Psi_{k_{\parallel}}(r)$\/. Since $k_{\text{F}}$\/ is increased as
compared to $k_{\text{F0}}$\/ when the barrier transmittance is
finite, the minimum in $\varepsilon_k$\/ is expected to occur at
larger $k_{\parallel}$\/.

With increasing the temperature $T$\/ the energy gap in the S region
$\Delta(T)$\/ is diminished, and the condition Eq.(\ref{limitcond}) is
violated at $T$\/ sufficiently close to $T_c$\/. The dependence of the
energy gap in the 2DEG $\varepsilon_g$\/ on $\Delta(T)$\/ is
determined from Eq.(\ref{dispersion2}) if we put
$(\xi_k-\frac{w}{s}\varepsilon_{g0})=0$\/, which means putting
$\varepsilon_k$\/ at a minimum. Then we obtain the equation for
$\varepsilon_g$
\begin{equation}
\varepsilon_g^2\left[ 1 + \frac{2\varepsilon_{g0}}{(\Delta(T)^2 -
\varepsilon_g^2)^{1/2}}\right]=\varepsilon_{g0}^2.
\label{gapEq}
\end{equation}
\noindent This dependence is shown in Fig.~4. The maximal value of
$\Delta(T)/\varepsilon_{g0}$\/ equals $\Delta(0)/\varepsilon_{g0}$\/. If
this value is very large, the energy gap in the excitation spectrum of the
2DEG coincides with $\varepsilon_{g0}$\/ in a wide range of $T$\/. The
characteristic temperature $T^{\ast}$\/ determining a transition of
$\varepsilon_g$\/ from $\varepsilon_{g0}$\/ to $\Delta(T)$\/ is given by
the equation $\Delta(T^{\ast})\cong \varepsilon_{g0}$\/. At low
temperatures ($\Delta(T)=\Delta(0)$\/) we can also calculate the influence
of the barrier, expressed in $w$\/, from Eq.(\ref{gapEq}). Fig.~5 shows the
energy gap $\varepsilon_g$\/ in the 2DEG as function of the
transmittance $T_b=1/((w/2s)^2+1)$\/ of the S-2DEG interface, for
different values of $s$\/.

If the excitation energy $\varepsilon_k$\/ exceeds $\Delta$\/, the wave
functions in the S region do not decay, but oscillate, and the $k$\/ vector
runs over a continuous set of values. These wave functions describe the
propagation of two electrons (incident on the barrier and reflected from
it) and a hole that appears as a result of Andreev reflection. The bound states
obtained above are closely related to those studied earlier in the 3D case
\cite{dGS-J63,B-SE-W77,Arn82}.

The wave functions corresponding to the bound states are determined by
Eq.(\ref{wavefunc}). By introducing new variables
\begin{equation}
\tilde{\varepsilon}_k\equiv\varepsilon_k/\varepsilon_0 =
\alpha_k+\xi_k/\varepsilon_0, ~t_k=\xi_k/\varepsilon_0-t_w,
{}~t_w=\frac{w}{w^2+s^2},
\label{newvar}
\end{equation}
we can write for the coefficients from Eq.(\ref{wavefunc}) the
relations
\begin{eqnarray}
E_-&=&E_+\exp(2i\alpha),~H_-=H_+\exp(2i\bar{\alpha})\nonumber\\
B_+&=&2\tilde{\varepsilon}_k\left[(E_+u_p +
H_+v_p)\tilde{\varepsilon}_k - (E_+u_p-v_pH_+)(t_k+t_w)\right] /
(\delta^2-\tilde{\varepsilon}_k)^{1/2}\nonumber\\
B_-&=&2\tilde{\varepsilon}_k\left[(E_+v_p +
H_+u_p)\tilde{\varepsilon}_k - (E_+v_p-u_pH_+)(t_k+t_w)\right] /
(\delta^2-\tilde{\varepsilon}_k)^{1/2}.
\label{coefficients}
\end{eqnarray}
\noindent Using these relations and Eq.(\ref{wavefunc}) for the wave
functions, we can find the Green's functions $\tilde{G}^{R(A)}$\/ for
the 2DEG in the system shown in Fig.1b and calculate the critical
Josephson current.

\section{The Green's functions and the DC Josephson effect}

Consider the system shown in Fig.~1b. Electrons move in the quantum well in
the $(x,y)$\/ plane. As shown before, the wave functions in the 2DEG change
drastically due to the proximity effect. In particular, the condensate
Green's functions $F^{R(A)}$\/ are induced in the quantum well, and
therefore the Josephson effect is possible in this system. In order to
determine the critical current, we need to know the functions $F^{R(A)}$\/,
that is, the nondiagonal elements of the matrix
$G_{\alpha\beta}^{R(A)}:~F^{R(A)}(z,z';q) = G_{12}^{R(A)}(z,z';q)$\/. We
are interested in the current $I$\/ averaged over the quantum well width,
which means that we must find $F^{R(A)}(k)=\langle
F^{R(A)}(z,z';q)\rangle$\/, where the brackets denote averaging over
$z~(0<z<d)$\/. As is well known, the matrix components of
$\tilde{G}^{R(A)}$\/ are expressed through the components of the wave
functions $\Psi_{\alpha}(z,q)$
\begin{equation}
G_{\alpha\beta}^{R(A)}(z,z;k)=\sum_{i=\pm 1}
\frac{\langle\Psi_{\alpha}(z,k)
\Psi_{\beta}^{\ast}(z,k)\rangle_i}{\varepsilon \pm i0 -
\varepsilon_{ki}},
\label{sumG}
\end{equation}
\noindent where the sum is taken over the two branches of the spectrum
determined by Eq.(\ref{dispersion2}). We suppose that the
transmittance of the SN barrier is small, implying
$\varepsilon_g\ll\Delta$\/, and the conditions $s/w\ll 1$\/ and
$s/w^2\ll\Delta/\varepsilon_0$\/ are fulfilled. Then, one can show
that $E_+$\/ and $H_+$\/ are coupled by the relation
\begin{equation}
E_+(\tilde{\varepsilon}_k-t_k)=\tilde{\varepsilon}_{g0}H_+.
\label{relEH}
\end{equation}
\noindent Taking into account Eqs.(\ref{relEH}) and
(\ref{lowTrlimit}), one can find from Eq.(\ref{sumG}) that the components
of the Green's functions are equal to
\begin{eqnarray}
G_{11}^{R(A)}(k) &=& \frac{1}{d} \left\{ \frac{1}{2\varepsilon_k}
\left[ \frac{\varepsilon_k+\varepsilon_0t_k}{\varepsilon\pm
i0-\varepsilon_k} +
\frac{\varepsilon_k-\varepsilon_0t_k}{\varepsilon\pm i0+\varepsilon_k}
\right] \right\}\nonumber\\ G_{12}^{R(A)}(k) &=& \frac{1}{d} \left\{
\frac{\varepsilon_{g0}}{2\varepsilon_k} \left[ \frac{1}{\varepsilon\pm
i0-\varepsilon_k} + \frac{1}{\varepsilon\pm i0+\varepsilon_k} \right]
\right\}.
\label{compG}
\end{eqnarray}
Here $\varepsilon_k=(\varepsilon_{g0}^2+\varepsilon_0^2t_k^2)^{1/2}$\/
is the excitation energy, $\varepsilon_g$\/ is the energy gap in the
excitation spectrum in the 2DEG at $|x|>L/2$\/ (see Eq.(\ref{gapEq})).
These functions are identical to the Green's functions of an ordinary
two dimensional superconductor with a spatially dependent energy gap
because, as supposed, the parameter $w(x)$\/ varies from a constant
value $w$\/ at $|x|>L/2$\/ to $\infty$\/ at $|x|<L/2$\/. The
characteristic length of the $w(x)$\/ variation is $x_0$\/, which is
small as compared to the coherence length in the 2DEG,
$\xi_{\text{2D}}=\hbar v_{\text{F}}/\pi\varepsilon_g$\/ and large as
compared to the Fermi wave length $k_{\text{F}}^{-1} =
(m_2v_{\text{F}})^{-1}$\/. Therefore the system under consideration is
equivalent to a 2DEG contacting at $|x|>L/2$\/ the superconducting
2DEG with the effective order parameter $\varepsilon_g\exp(\pm
i\phi/2)$\/, where $\varepsilon_g\ll\Delta$\/ (strictly speaking this
magnitude is achieved at $|x|>L/2$\/ over distances of the order of
the coherence length as it takes place in ordinary SNS junctions).
Hence for the critical current $I_c$ one can use the formulae obtained
in Refs.~\onlinecite{Kul70,Ish70,BJ72} where the width of the N layer
was assumed to be large or in Ref.~\onlinecite{BvH91b} where a one
dimensional channel is analyzed. If the width, $L_y$\/, of a channel
in the 2DEG shown in Fig.~1b is comparable with the Fermi wave length
$k_{\text{F}}^{-1}$\/ at $|x|<L/2$\/ and is much larger at
$|x|>L/2$\/, we can use the expression for $I_c$\/ obtained in
Ref.~\onlinecite{BvH91b}
\begin{equation}
I_c = N (e\varepsilon_g/\hbar) \sin(\phi/2)
\tanh((\varepsilon_g/2T)\cos(\phi/2)),
\label{Ic}
\end{equation}
\noindent where $N$\/ is the number of subbands below the Fermi level.
This means that $I_c$\/ will increase step wise with increasing the
electron density in the quantum well, and the height of the steps
equal $e\varepsilon_g/\hbar$\/.

\section{Conclusions}

We have found the critical Josephson current $I_c$\/ for a 2DEG in contact
with two superconductors. In contrast to previously analyzed systems
the effective order parameter in the 2DEG $\varepsilon_g\exp(i\phi)$\/
is reduced in comparison with $\Delta\exp(i\phi)$\/ in the
superconductor, and its magnitude
is determined by the SN barrier (Schottky barrier) transmittance and
by a mismatch of the Fermi momenta and the effective masses in the S
and N regions (here $\phi$\/ is the macroscopic phase of the
superconductors). The barrier transmittance characterized by the
parameter $w$\/ may depend on the carrier density in the 2DEG,
$n_{\text{2D}}$\/. Therefore, the critical current $I_c$\/ will depend
on $n_{\text{2D}}$\/ even when only the lowest subband is filled.

One of the authors (AFV) is grateful to the Materials Science Center of the
University of Groningen, where a part of this work was performed, for
hospitality and useful discussions. This work was supported financially by
the NATO Linkage Grant \# 921168 and by the Russian Fund for Fundamental
Research (project \# 93 02 15042). The work at the University of Groningen
is supported by the Stichting voor Fundamenteel Onderzoek de Materie (PHCM)
and by the Royal Academy of Sciences of the Netherlands (BJvW).

\begin{figure}
\caption{Schematic representation of S-2DEG-S Josephson junctions, as
studied in Refs.~{\protect\onlinecite{BvH91b,FTT91}} (1a) and in the
present paper (1b). The spacing between the superconductors is $L$\/,
the width of the 2DEG is $d$\/.}
\end{figure}

\begin{figure}
\caption{Energy diagram of the system under consideration -- a
superconductor and a 2DEG are separated by a potential barrier
$U_b$\/. Hatched areas denote states filled with electrons (we suppose
that only the lowest subband in the quantum well, with energy
$\varepsilon_0$\/, is occupied). The energy gap in the excitation
spectrum of the 2DEG is induced due to the presence of the
superconductor.}
\end{figure}

\begin{figure}
\caption{The normalized excitation energy in the 2DEG,
$\varepsilon_k/\Delta$\/, vs the longitudinal momentum of electrons,
$k_{\parallel}$\/, at different values of the parameter
$\varepsilon_{g0}/\Delta = (\varepsilon_0/\Delta)s/(s^2+w^2)$\/:
$\varepsilon_{g0}/\Delta = 0.2$\/ (dashed line); 0.3 (solid line); 0.6
(dotted line).}
\end{figure}

\begin{figure}
\caption{The normalized energy gap in the excitation spectrum of the
2DEG $\varepsilon_g/\varepsilon_{g0}$\/ vs the normalized energy gap
in the S region $\Delta/\varepsilon_{g0}$\/. When the temperature is
increased towards $T_c$\/, $\Delta(T)$\/ goes to 0, thus violating
the condition of Eq.({\protect\ref{limitcond}}).}
\end{figure}

\begin{figure}
\caption{The normalized energy gap $\varepsilon_g/\Delta$\/ vs the
transparency $T_b$\/, expressed in the dimensionless parameter $w$\/,
of the barrier between superconductor and 2DEG, for different values
of the mismatch, $s=p_{\text{F}}m_2/k_0m_1$\/: $s=0.5$\/
(solid line); 1 (dashed line); 2 (dotted line).}
\end{figure}


\begin{thebibliography}{99}

\bibitem{NATA92}
J. Nitta, T. Akazaki, H. Takayanagi, and K. Arai, Phys. Rev. B {\bf
46}, 1486 (1992).

\bibitem{NKH92}
C. Nguyen, H. Kroemer, and E.~L. Hu, Phys. Rev. Lett. {\bf 69}, 2847
(1992).

\bibitem{DHvWetal94}
A. Dimoulas {\em et al\/}, submitted to Phys. Rev. Lett., 1994.

\bibitem{Letal93}
K.~M.~H. Lenssen {\em et al.\/}, Proc. of Applied Superconductivity
Conference, 1993 (Chicago) IEEE Trans., MAG-27, 1993.

\bibitem{Getal93}
J.~R. Gao {\em et al.\/}, Report presented at 10th Intern. Conf. on
Electronic Properties of Two-Dimensional Systems, Newport, Rhode
Islands, USA, 1993.

\bibitem{MWA94}
A.~M. Marsh, D.~A. Williams, and H.~Ahmed, To be published in Phys. Rev. B,
Rap. Comm., 1994.

\bibitem{vWetal88}
B.~J. van Wees {\em et al.\/}, Phys. Rev. Lett. {\bf 60}, 848 (1988).

\bibitem{Wetal88}
D.~A. Wharam {\em et al.\/}, J. Phys. C {\bf 21}, L209 (1988).

\bibitem{BvH91a}
C.~W.~J. Beenakker and H. van Houten, in Solid State Physics, ed. by
H. Ehrenreich and D. Turnbull (Academic, New York, 1991), Vol.~{\bf
44}, p.~1.

\bibitem{BvH91b}
C.~W.~J. Beenakker and H. van Houten, Phys. Rev. Lett. {\bf 66}, 3056
(1991)

\bibitem{FTT91}
A. Furusaki, H. Takayanagi, and M. Tsukada, Phys. Rev. Lett. {\bf 67},
132 (1991); Phys. Rev. B {\bf 45}, 10563 (1992).

\bibitem{Kul70}
I.~O. Kulik, Sov. Phys. JETP {\bf 30}, 944 (1970).

\bibitem{Ish70}
C. Ischii, Prog. Theor. Phys. {\bf 44}, 1525 (1970).

\bibitem{BJ72}
J. Bardeen, and J.~L. Johnson, Phys. Rev. B {\bf 5}, 72 (1972).

\bibitem{McM68}
W.~L. McMillan, Phys. Rev. {\bf 175}, 537 (1968).

\bibitem{MB84}
R.~A. Morrow, and K.~R. Brownstein, Phys. Rev. B {\bf 30}, 678 (1984).

\bibitem{dGS-J63}
P.~G. de Gennes, and D. Sain-James, Phys. Lett. {\bf 4}, 151 (1963).

\bibitem{B-SE-W77}
J. Bar-Sagi, and O. Entin-Wohlman, Solid State Comm. {\bf 22}, 29
(1977).

\bibitem{Arn82}
G. Arnold, Phys. Rev. B {\bf 25}, 5998 (1982); {\bf 23}, 1171 (1981);
{\bf 18}, 1076 (1978).

\end{thebibliography}
\end{document}